\documentclass{article}

\usepackage{apj}
\lefthead{Aloy et al.}
\righthead{RELATIVISTIC JETS FROM COLLAPSARS}

\def\be{\begin{equation}}
\def\ee{\end{equation}}
\newcommand\cref[1]{(\ref{#1})}

 \def\etal{{\sl et al.}\,}
 \def\ie{{\sl i.e.,}\,}
 \def\eg{{\sl e.g.,}\,}

 \def\la{\hbox{\raise.5ex\hbox{$<$}
     \kern-1.1em\lower.5ex\hbox{$\sim$}}}
 \def\ga{\hbox{\raise.5ex\hbox{$>$}
     \kern-1.1em\lower.5ex\hbox{$\sim$}}}

 \def\sun{\odot}
 \def\ms{M_{\sun}}

\begin{document}

\title{RELATIVISTIC JETS FROM COLLAPSARS}

\author{M.A. Aloy\altaffilmark{1}, 
        E. M\"uller\altaffilmark{2},
        J.M$^{\underline{\mbox{a}}}$ Ib\'a\~nez\altaffilmark{1},
        J.M$^{\underline{\mbox{a}}}$ Mart\'{\i}\altaffilmark{1} and
        A. MacFadyen\altaffilmark{3}
}

\altaffiltext{1} {Departamento de Astronom\'{\i}a y Astrof\'{\i}sica,
                  Universidad de Valencia, 46100 Burjassot (Valencia), Spain}

\altaffiltext{2} {Max-Planck-Institut f\"ur Astrophysik,
                  Karl-Schwarzschild-Str. 1, 85748 Garching, Germany}

\altaffiltext{3} {Astronomy Department, University of California, 
                  Santa Cruz, CA 95064}

\begin{abstract}
Using a collapsar progenitor model of MacFadyen \& Woosley we have
simulated the propagation of an axisymmetric jet through a collapsing
rotating massive star with the GENESIS multi-dimensional relativistic
hydrodynamic code. The jet forms as a consequence of an assumed
(constant or variable) energy deposition in the range
$10^{50}$\,erg\,s$^{-1}$ to $10^{51}$\,erg\,s$^{-1}$ within a
$30^{\circ}$ cone around the rotation axis. The jet flow is strongly
beamed ($\la$ few degrees), spatially inhomogeneous, and time
dependent. The jet reaches the surface of the stellar progenitor
($R_{\ast} = 2.98 \times 10^{10}\,$cm) intact.  At breakout the
maximum Lorentz factor of the jet flow is 33.  After breakout the jet
accelerates into the circumstellar medium, whose density is assumed to
decrease exponentially and then being constant $\rho_{\rm ext} =
10^{-5}$\,gcm$^{-3}$. Outside the star the flow begins to expand also
laterally ($v \sim c$), but the beam remains very well collimated. At
a distance of $2.54\,R_{\ast}$, where the simulation ends, the Lorentz
factor has increased to 44.
\end{abstract}


\section{INTRODUCTION } \label{s:intro}

Catastrophic collapse events have been proposed to explain the
energies released in a gamma-ray burst (GRB) including mergers of
compact binaries (Pacy\'nski 1986; Goodman 1986; Eichler \etal 1989;
Mochkovitch \etal 1993), collapsars (Woosley 1993) and hypernovae
(Pacy\'nski 1998). According to the current view these models require
a stellar mass black hole (BH) which accretes up to several solar
masses of matter powering a pair fireball. If the baryon load of the
fireball is not too large, baryons are accelerated together with
e$^+\,$e$^-$ pairs to Lorentz factors $> 10^2$ (Cavallo \& Rees
1978). Such relativistic flows are supported by radio observations of
GRB\,980425 (Kulkarni \etal 1998b).
Spherically symmetric fireballs have been studied by several authors
by means of 1D Lagrangian hydrodynamic simulations (\eg Panaitescu
\etal 1997; Panaitescu \& M\'esz\'aros 1998; Kobayashi, Piran \& Sari
1999).  Recently, it has been argued that the rapid temporal decay of
several GRB afterglows is more consistent with the evolution of a
relativistic jet after it slows down and spreads laterally than with a
spherical blast wave (Sari, Piran \& Halpern 1999; Halpern \etal 1999;
Kulkarni \etal 1999; Rhoads 1999). The lack of a radio afterglow in
GRB\,990123 provides independent evidence for jet-like geometry
(Kulkarni \etal 1999b).

\section{INITIAL MODEL AND NUMERICAL SETUP} 
\label{s:setup}

MacFadyen \& Woosley (1999; MW99) have explored the evolution of
rotating helium stars ($M_{\alpha}\, \ga\, 10\,\ms$) whose iron core
collapse does not produce a successful outgoing shock, but instead
forms a BH surrounded by a compact accretion torus.  Assuming an
enhanced efficiency of energy deposition in polar regions MW99 obtain
relativistic jets along the rotation axis, which are highly focused
and seem to be capable of penetrating the star.  However, as their
simulations are Newtonian, they obtain flow speeds which are
superluminal.

We have performed axisymmetric relativistic simulations using a
$14\,\ms$ collapsar model from MW99. When the central BH has acquired
a mass of $3.762\,\ms$ we map the model to our computational grid.  In
a consistent collapsar model a jet will be launched by any process
which gives rise to a local deposition of energy and/or momentum, as
\eg $\nu \bar\nu$--annihilation, or magneto-hydrodynamic processes. We
mimic such a process by depositing energy at a prescribed rate
homogeneously within a $30^{\circ}$ cone around the rotation axis.  In
the radial direction the deposition region extends from the inner grid
boundary located at 200\,km to a radius of 600\,km.  We have
investigated constant energy deposition rates $\dot E =
10^{50}\,$erg\,s$^{-1}$, and $\dot E = 10^{51}\,$erg\,s$^{-1}$, and a
varying deposition rate with a mean value of
$10^{50}\,$erg\,s$^{-1}$. The constant rates roughly bracket the
expected $\dot E$ of collapsar models, while the varying rate mimics,
\eg time-dependent mass accretion rates resulting in time-dependent
$\nu \bar \nu $--annihilation (MW99).

The simulations were performed with the multidimensional relativistic
hydrodynamic code GENESIS (Aloy \etal 1999) using a 2D spherical grid
with 200 radial zones spaced logarithmically between the inner
boundary and the surface of the helium star at $R_{\ast} = 2.98 \times
10^{10}\,$cm. Assuming equatorial symmetry we performed simulations
with $2^{\circ}, 1^{\circ}$ and $0.5^{\circ}$ angular resolution. In
the latter case the grid consists of 60 uniform zones covering the
polar region ($0^{\circ} \le \theta \le 30^{\circ}$) and 40 nonuniform
zones logarithmically distributed between $30^{\circ} \le \theta \le
90^{\circ}$.  The gravitational field of the BH is described by the
Schwarzschild metric.  Effects due to the self-gravity of the star are
neglected, \ie we consider only the gravitational potential of the BH.
The equation of state includes non-relativistic nucleons treated as a
mixture of Boltzmann gases, radiation, and an approximate correction
due to $e^+e^-$--pairs as described in Witti, Janka \& Takahashi
(1994).  Complete ionization is assumed, and the effects due to
degeneracy are neglected. We advect nine non-reacting nuclear species
which are present in the initial model: C$^{12}$, O$^{16}$, Ne$^{20}$,
Mg$^{24}$, Si$^{28}$, Ni$^{56}$, He$^{4}$, neutrons and protons.

\section{RESULTS}  
\label{s:results} 

\subsection{Constant small energy deposition rate (Model C50)}
For a constant $\dot E = 10^{50}\,$erg\,s$^{-1}$ a relativistic jet
forms within a fraction of a second and starts to propagate along the
rotation axis with a mean speed of $7.8 \times 10^{9}$\,cm\,s$^{-1}$
(Fig.\,1).  The jet exhibits all the morphological elements of the
Blandford \& Rees (1974) jet model in the context of classical double
radio sources: a terminal bow shock, a narrow cocoon, a contact
discontinuity separating stellar and jet matter, and a hot spot.
Fig.\,1 shows that the density structure of the star does not change
noticeably during the whole evolution.  This, a posteriori, justifies
our neglect of the self-gravity of the star.

The propagation of the jet is unsteady, because of density
inhomogeneities in the star.  The Lorentz factor of the jet, $\Gamma$,
increases non-monotonically with time, while the density drops to
$\sim 10^{-6}\,$g\,cm$^{-3}$ (Fig.\,2). The density profile shows
large variations (up to a factor of 100) due to internal shock waves.
The mean density in the jet is $\sim 10^{-1}$\,g\,cm$^{-3}$.  Some of
the internal biconical shocks, which develop during the jet's
propagation, recollimate the beam.  They may provide the ``internal
shocks'' proposed to explain the observed gamma-ray emission.  A
particularly strong recollimation shock wave (hardly evident at low
resolution) forms early in the evolution. A very strong rarefaction
wave behind this recollimation shock causes the largest local
acceleration of the beam material giving rise to a maximum in the
Lorentz factor.  When the jet encounters a region along the axis where
the density gradient is positive (at $\log r \approx 8.1$ and $\log r
\approx 8.6$) the jet's head is decelerated, while a central channel
in the beam is cleaned by outflow into the cocoon through the head,
which accelerates the beam. The combination of both effects
(deceleration of the head and beam acceleration) increases the
strength of the internal shocks. Within the jet the mean value of the
specific internal energy, $\sim 10^{20} - 10^{21}$\,erg\,g$^{-1}$, or
$\sim O(c^2)$.  The mean temperature $\sim 5\times10^8$\,K (well below
the pair creation threshold) implying that the pressure is radiation
dominated in accordance with our simplified EOS.

The relativistic treatment of the hydrodynamics leads to a
qualitatively similar (formation of a jet), but quantitatively very
different evolution than in MW99. According to their Fig.\,27 the jet
propagates 7\,000\,km within the first 0.82\,s. Furthermore, MW99
infer an asymptotic $\Gamma \sim 10$, and find a half opening angle,
$\Omega$, for their jet of $\sim 10^{\circ}$. In our simulation, at
the same time for the same angular resolution ($\sim 2^{\circ}$) and
$\dot E$ the head reaches a radius of 30\,000\,km, but the maximum
Lorentz factor ($\Gamma_{\rm{max}}$) is only 4.62 at $\sim
12\,200$\,km.  Such quantitative difference is expected, in part, due
to different mapping time and inner boundary radius between the two
calculations.  Initially, in our simulations $\Omega$ is between
$6^{\circ}$ to $8^{\circ}$ depending on angular resolution.  At $\sim
1.5$s the strong recollimation shock reduces $\Omega \la 1^{\circ}$.

We find that some results strongly depend on angular resolution, the
minimum acceptable one being $0.5^{\circ}$ (at least near the axis).
The morphology of the jet is richer at higher resolution.  At
$0.5^{\circ}$ angular resolution $\Gamma_{\rm{max}} \approx 15-20$ at
a radius $\sim 8 \times 10^9$\,cm at jet breakout.  Within the
uncertainties of the jet mass determination due to finite zoning and
the lack of a precise numerical criterion to identify jet matter, the
baryon load ($\eta \equiv M c^2/ E_{\rm depos}$ with $E_{\rm
depos}=\int \dot{E} dt$) decreases with increasing resolution. In the
highest resolution run we find an average baryon load of ${\eta}
\simeq 1.3$ at jet breakout (see also Sect.\,4).

\subsection{Constant large energy deposition rate (Model C51)}

Enhancing $\dot E$ by a factor of ten (to $10^{51}$\,erg\,s$^{-1}$),
the jet flow reaches larger Lorentz factors. We observe transients
during which the Lorentz factor becomes as large as 40.  After 1.2\,s
the Lorentz factor steadily increases from 22 to 33. The jet
propagates faster than in model C50.  The time required to reach the
surface of the star is 2.27\,s instead of 3.35\,s.  At breakout the
jet is less collimated ($\Omega \sim 10^{\circ}$).  The strong
recollimation shock present in model C50 is not so evident
here. Instead, several biconical shocks are observed within a very
knotty beam and the Lorentz factor near the head of the jet is larger
($\sim 22$ in the final model) because, due to the larger $\dot E$,
the central funnel is evacuated faster, and because the mean density
of the jet is 5 times smaller than in model C50 ($\eta$ being twice as
large).

\subsection{Varying energy deposition rate (Model V50)}

We have computed a model where the mean energy deposition rate
($10^{50}$\,erg\,s$^{-1}$) randomly varies on time scales of a few
milliseconds and the amplitude by a factor of ten.  Compared to model
C50 the jet structure is more knotty and also richer in shocks,
particularly inside the first $10^9$\,cm (where the radial resolution
is large enough to capture the finest structures imprinted on
the flow by the time variability of the deposition rate). At breakout
$\Gamma_{\rm max} = 26.81$, which is {\it almost twice} as large as
the one found in model C50.  Thus, a variable $\dot E$ is more
efficient in converting internal energy into kinetic energy, and
in this case, the internal shocks are stronger and more numerous.  The
mean propagation speed is similar in both models, although the
instantaneous velocity of the jet's head is clearly different. Behind
the strongest recollimation shock $\Omega < 1^{\circ}$ in both models.

\subsection{Evolution after jet breakout}

The structure of the circumstellar medium will influence the
characteristics of the GRB and of the subsequent afterglow. Thus, a
continuation of the simulations beyond jet breakout is necessary.  In
order to satisfy the conditions for accelerating shocks (Shapiro 1979)
we endowed the star with a Gaussian atmosphere, which at $R_{\rm a} =
1.8\, R_{\ast}$ passes over into an external uniform medium with a
density $10^{-5}$\,g\,cm$^{-3}$ and a pressure $10^{-8}
p(R_{\ast})$. The computational domain is extended to $R_t = 2.54
R_{\ast}$ with 70 additional zones.  The evolution after jet breakout
has been computed for models C50 and C51. In both cases the jet
reaches $R_t$ after $\sim 1.8$\,s (measured from breakout). Its mean
propagation velocity is $\sim 0.85$\,c, which is almost three times
faster than the velocity of the head inside the star (0.30\,c in model
C50; 0.44\,c in model C51).

The evolution after jet breakout consists of three distinct epochs
(Fig.\,3).  The first one lasting 0.35\,s is characterized by a head
velocity of 0.48\,c and a small sideways expansion.  During the second
phase (of 0.3\,s) the jet head accelerates to 0.91\,c, because of the
steep external density gradient, and because the flux of axial
momentum is still important compared to pressure.  The sideways
expansion is still sub-relativistic ($\approx 0.008$\,c), and $\Omega$
of the beam increases to $\approx 10^{\circ}$.  During the final
1.2\,s the bow-shock propagates within the uniform part of the ambient
medium leading to a rapid ($\Gamma \simeq 5$) lateral spreading
(Fig.\,3).

The shape of the expanding bubble is prolate (Figs.\,1 and 3) during
the post-breakout evolution.  However, when the jet reaches the
uniform part of the circumstellar environment, the bubble widens due
to the faster sideways expansion. We expect a more isotropic expansion
when most of the bubble is inside the uniform medium, and when it is
pressure driven (in particular if the energy deposition is switched
off).  The Lorentz factor near the boundary of the cavity blown by the
jet grows from $\sim 1$ (at jet breakout) to $\sim 3$ in both models
decreasing with latitude.  At the end of the simulation $\Gamma_{\rm
max}$ is 29.35 (44.17) for model C50 (C51), which is still smaller
than the ones required for the fireball model (Piran 1999).

\section{DISCUSSION AND CONCLUSIONS}

Energy deposition in the polar regions of a collapsar model gives rise
to both the formation and propagation of a relativistic jet through
the mantle and envelope of the star, and to a supernova explosion. The
jet has a small opening angle ($\sim 8^\circ$) and possesses a highly
collimated ($ \sim 1^\circ$), ultra-relativistic core in which the
Lorentz factor reaches a value of $\Gamma_{\rm{max}} = 44$ (model C51)
at the end of the simulation about 2\,s after shock breakout. The
equivalent isotropic kinetic energy (see MacFadyen, Woosley \& Heger
1999) slightly exceeds $10^{54}$\,erg for model C51 ($10^{53}$\,erg
for model C50) within $2^\circ$ ($5^\circ$) of the rotation axis
dropping by a factor of 10 within $17^\circ$ ($10^\circ$).  The inner
region contains $8\times 10^{-4}\,\ms$ with $< \Gamma > \sim 4$
(Fig.\,1).  For a larger $\dot E$, the jet and in particular the
cocoon are less collimated, because a harder driven jet also expands
stronger laterally.

The rest-mass density and the internal energy strongly vary in space
and time within the jet giving rise to a very inhomogeneous baryon
load, \ie the concept of ${\eta}$ as a global parameter is
useless. Instead it is more appropriate to discuss the efficiency of
energy conversion in terms of incremental baryon loads considering
only matter within a given range of $\Gamma$--values.  Although we
find an average baryon load of the jet of $\bar{\eta} \sim 1$, some
parts of the flow have a baryon load as low as $\sim 10^{-5}$ or even
less.  After jet breakout $\bar{\eta}$ decreases by a factor 4 in less
than 1.8\,s. If this trend continues even $\bar{\eta} \sim 10^{-3}$
within 9\,s.

In model C51, at the end of the simulation, $\sim 2\,\ms$ have a
Lorentz factor of less than three, $3\times 10^{-4}\,\ms$ move with $3
\le \Gamma < 10$, and for $2\times 10^{-6}\,\ms$ the Lorentz factor
$\Gamma \ge 10$ (Fig.\,4). The latter two masses reduce to $2\times
10^{-5}\,\ms$ and $2\times 10^{-7}\,\ms$ for model C50.  Except for
the very early evolution ($t< 1$\,s) the amount of matter moving at
moderate ($3 \le \Gamma < 10$) and highly ($\Gamma \ge 10$)
relativistic velocities increases by a factor $\sim 3$ every second,
\ie if the central engine is active for another 5\,s, at the assumed
energy deposition rate, $\sim 10^{-4}\,\ms$ will move with $\Gamma \ge
10$. As $\Gamma_{\rm max}$ is also rapidly increasing, Lorentz factors
of several hundreds might be reached before the central engine is
switched off.  For models where the released total energy is equal
(C50 and V50) $\Gamma_{\rm max}$ is higher (by a factor of two) for a
time-dependent $\dot E$.  Determining the efficiency of energy
conversion ${\cal E} \equiv E_k / E_{\rm depos}$ is hampered by the
fact, that the kinetic energy $E_k$ of a relativistic fluid is not a
well defined quantity. If $E_k \equiv \int \rho \Gamma (\Gamma - 1)
dV$, we find ${\cal E} = 1.6$ for model C50 (and ${\cal E} = 2.8$ for
model C51) at the end of the simulation. These efficiencies are
obtained considering only matter with radial velocities $>0.3c$ and
specific internal energy densities $>5 \times 10^{19}$\,ergg$^{-1}$,
\ie matter in the jet. Note that efficiencies larger than one can
arise, because (i) there are also other large sources of energy (\eg
gravitational, internal) available, and (ii) because matter is
entrained into the jet which does not originate from the deposition
region. Thus, efficiencies larger than one suggest that the local
energy deposition is efficient triggering conversion of other forms of
energy into kinetic energy.

In our simulations the jet reaches the stellar surface intact
(propagating over three decades in radius). This result may also hold
for other less specific initial conditions. A more spherical density
stratification might decrease the collimation of the jet, but the
outflow might also be initiated mostly by momentum deposition instead
of pure energy deposition (\eg by MHD effects).  The propagation of
the jet after breakout will depend on the density stratification of
the circumstellar medium. Thus, further simulations with different
environments are planned. We note in this respect that the
post--breakout propagation is similar in models C50 and C51 suggesting
that a lower value of $\rho_{\rm ext}$ will not change the dynamics
quantitatively.  In our models the jet has only reached a radius of
$7.5 \times 10^{10}$\,cm at the end of our simulations, which is
$10^2$ to $10^4$ times smaller than the distance at which the fireball
becomes optically thin. Determining whether a GRB will eventually be
produced requires to compute the further evolution of the jet. As the
jet has stayed collimated in the star it might remain focused
over the next three decades, too.

\acknowledgements

This work has been supported in part by the Spanish DGES (grant
PB97-1432) and the CSIC. MAA expresses his gratitude to the
Conselleria d'Educaci\'o i Ci\`encia de la Generalitat Valenciana for
a fellowship. We would like to thank Stan Woosley for his enthusiasm
in promoting this work, and Thomas Janka for many helpful remarks.
The calculations were performed on two SGI Origin 2000 at the CEPBA
and at the SIUV.

\clearpage

\begin{figure}
\plotone{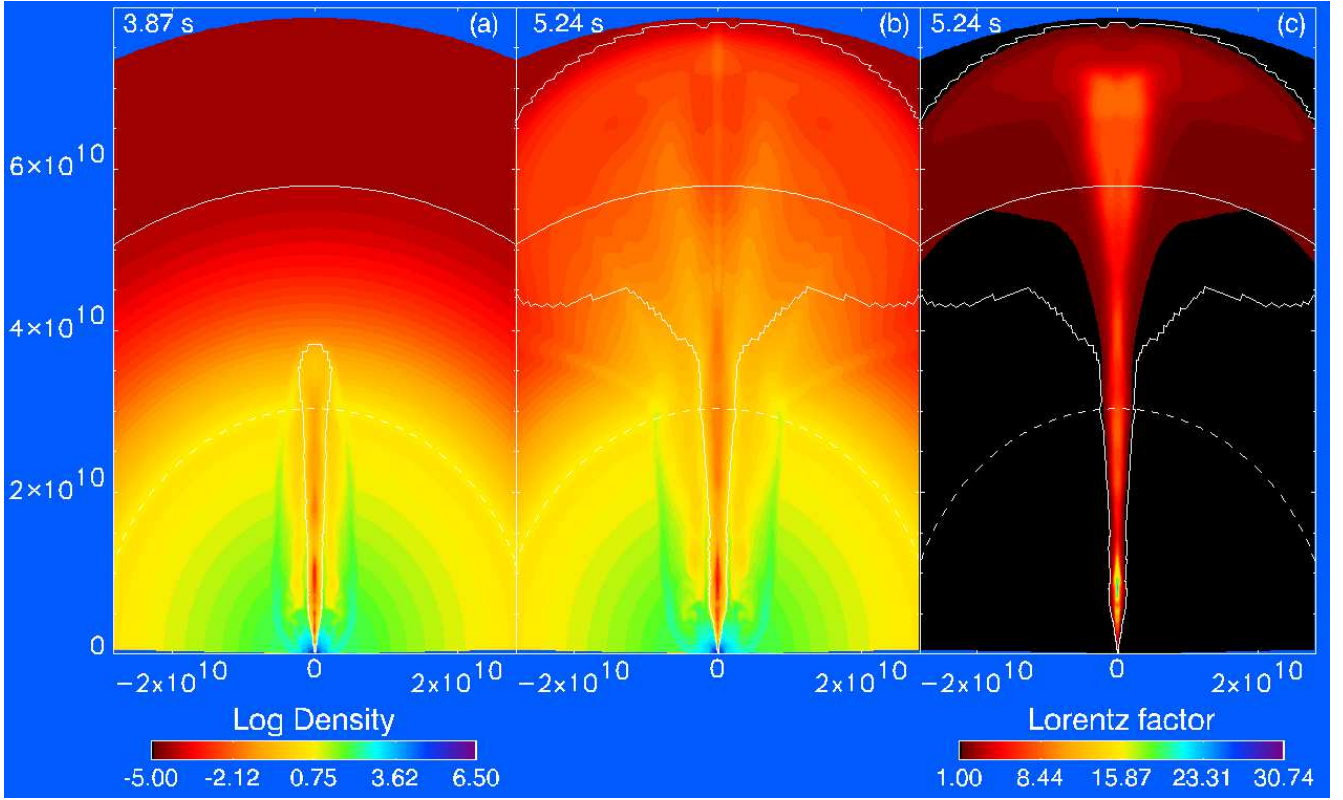}
\caption{Contour maps of the logarithm of the rest--mass density after
3.87\,s and 5.24\,s (left two panels), and of the Lorentz factor
(right panel) after 5.24\,s. X and Y axis measure distance in
centimeters. Dashed and solid arcs mark the stellar surface and
the outer edge of the exponential atmosphere, respectively. The other
solid line encloses matter whose radial velocity $>0.3c$, and
whose specific internal energy density $>5 \times
10^{19}$\,erg\,g$^{-1}$.
\label{f:loredens}}
\end{figure}

\begin{figure*}
\plotone{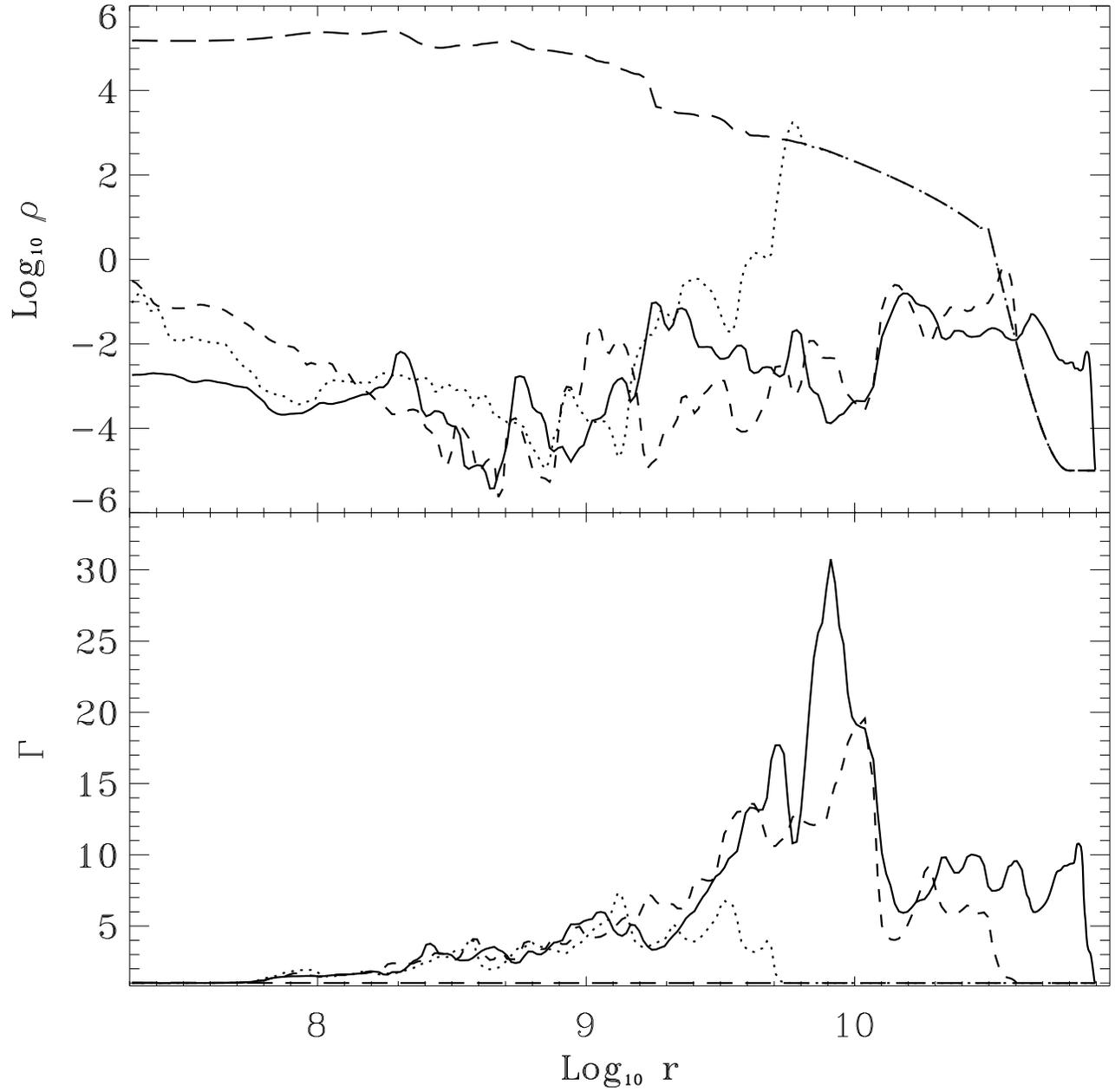}
\caption{Rest--mass density (top) and Lorentz factor (bottom) vs radius
along the symmetry axis for model C50 at $t=0$\,s (long dashed), $t=1.44$\,s
(dotted), $t=3.87$\,s (dashed) and $t=5.24$\,s (solid), respectively.
\label{f:polar_lor_den}}
\end{figure*}

\begin{figure*}
\plotone{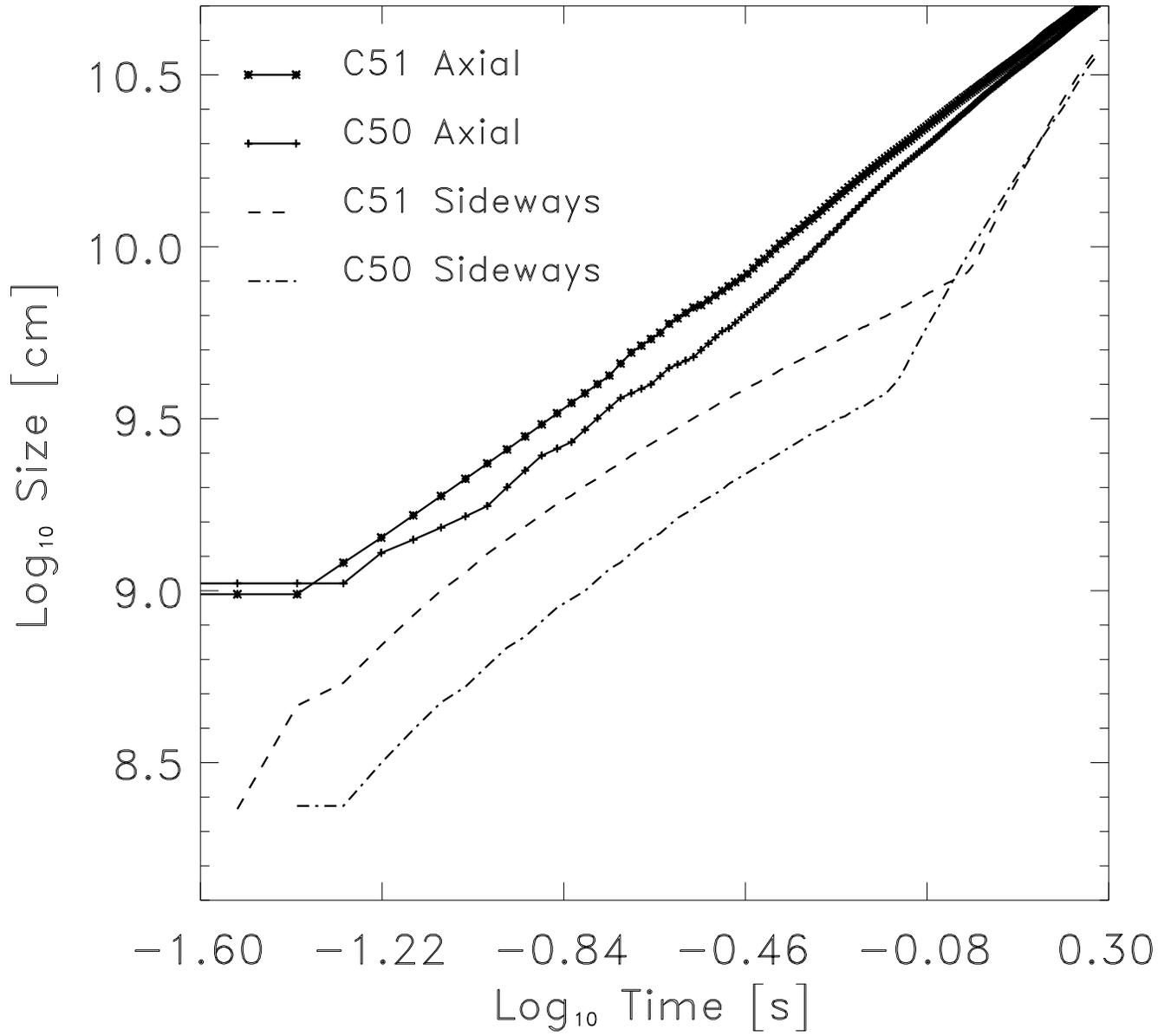}
\caption{Axial and lateral size of the jet driven bubble versus time
since jet breakout for models C50 and C51.
\label{f:size}}
\end{figure*}

\begin{figure*}
\plotone{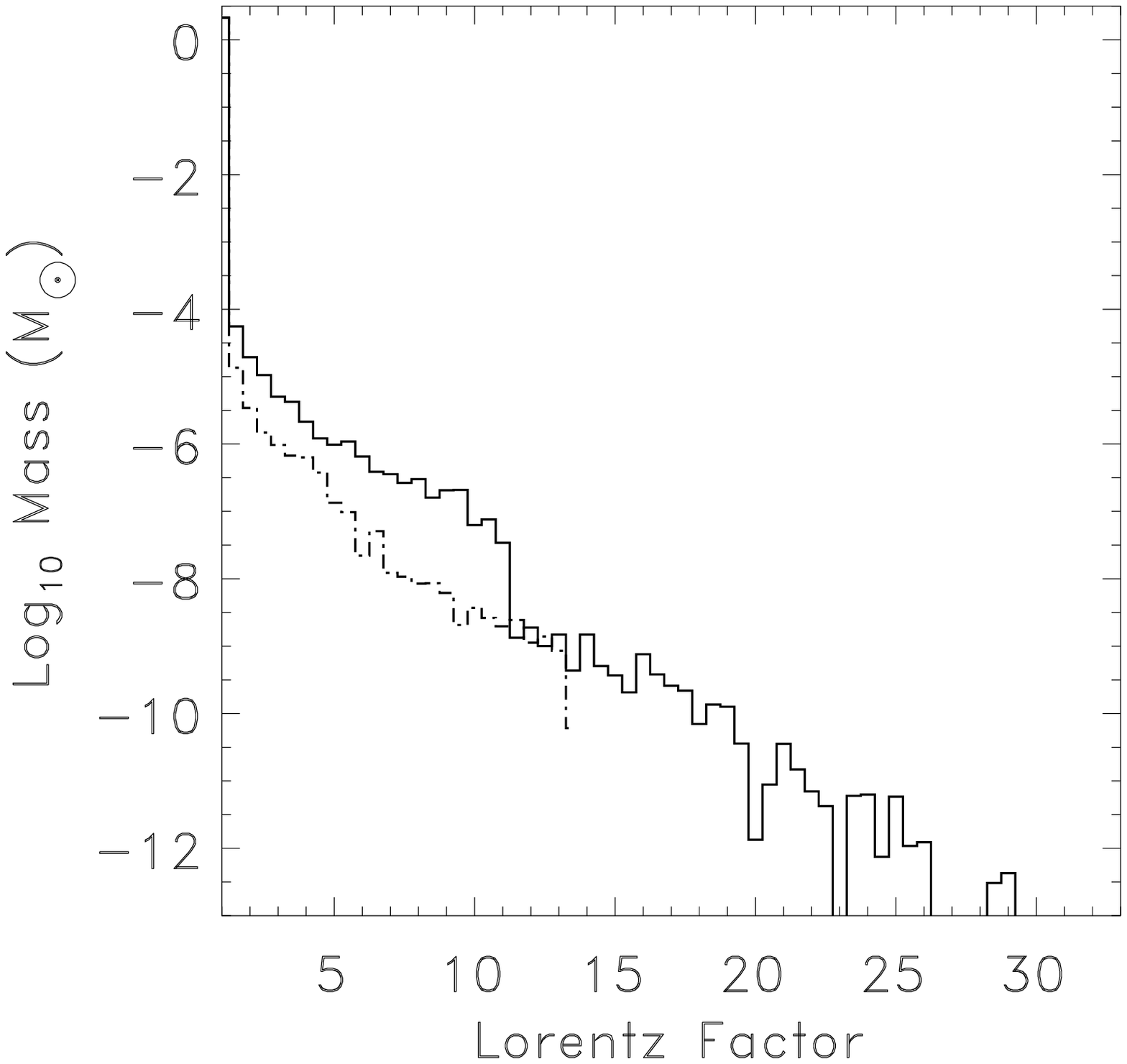}
\caption{Mass (binned in $\Delta \Gamma = 0.5$ intervals) versus Lorentz
factor for model C50 at shock breakout (dash-dotted) and at
$t=5.24$\,s (solid), respectively.
\label{f:loremass}}
\end{figure*}


\begin{thebibliography}{}

  \bibitem{AI99} Aloy, M.A., Ib\'a\~nez, J.M$^{\underline{\mbox{a}}}$,
                 Mart\'{\i}, J.M$^{\underline{\mbox{a}}}$ and M\"uller,
                 E. (1999), \apjs{122}, 151

  \bibitem{BR74} Blandford, R. and Rees, M.J. (1974), \mnras{169}, 395

  \bibitem{CR78} Cavallo, G. and Rees, M.J. (1978), \mnras{183}, 359

  \bibitem{Ei89} Eichler, D., Livio, M., Piran, T. and Schramm, D.N., 1989,
                 \nat{340}, 126

  \bibitem{Go86} Goodman, J. (1986), \apj{308}, L47
 
  \bibitem{HK99} Halpern, J.P, Kemp, J., Piran, T. and Bershady, M.A.
                 (1999), \apjl{517}, L105

  \bibitem{KP99} Kobayashi, S., Piran, T. and Sari, R. (1999),
                 \apj{513}, 669

  \bibitem{KF98} Kulkarni, S.R., Frail, D.A., Wieringa, M.H., Ekers,
                 R.D., Sadler, E.M., Wark, R.M., Higdon, J.L., 
                 Phinney, E.S. and Bloom, J.S. (1998b), \nat{395}, 663

  \bibitem{KD99} Kulkarni, S.R., Djorgovski, S.G., Odewahn, S.C.,
                 Bloom, J.S., Gal, R.R., \etal (1999a), \nat{398}, 389

  \bibitem{KF99} Kulkarni, S.R., Frail, D.A., Sari, R.,
                 Moriarty-Schieven, G.H., Shepherd, D.S., Udomprasert,
                 P., Readhead, A.C.S., Bloom, J.S., Feroci, M. and
                 Costa, E. (1999b), preprint astro-ph/9903441

  \bibitem{MW99} MacFadyen, A.I. and Woosley, S.E. (1999), 
                 \apj{524}, 262 (MW99)

  \bibitem{MF99} MacFadyen, A.I., Woosley, S.E., and  Heger, A. (1999), 
                 submitted to ApJ and astro-ph/9910034

  \bibitem{Me99} M\`esz\'aros, P. (1999), in {\it Proc. of the 19th Texas
                 Symp. on Relativistic Astrophysics \& Cosmology}, 
                 Paris, Dec.\,1998; and astro-ph/9904038

  \bibitem{MH93} Mochkovitch, R., Hernanz, M., Isern, J. and Martin, X.
                 (1993), \nat{361}, 236

  \bibitem{Pa86} Pacy\'nski, B. (1986), \apj{308}, L43

  \bibitem{Pa98} Pacy\'nski, B. (1998), \apj{494}, L45

  \bibitem{PW97} Panaitescu, A., Wen, L., Laguna, P. and M\'esz\'aros,
                 P. (1997), \apj{482}, 942

  \bibitem{PM99} Panaitescu, A. and M\'esz\'aros, P. (1999), \apj{}, 
                 in press; and astro-ph/9810258

  \bibitem{PA99} Pian, E., Amati, L., Antonelli, L.A., Butler, R.C., 
                 Costa, E., \etal (1999), \aap{}, submitted; and 
                 astro-ph/9903113

  \bibitem{Pi99} Piran, T. (1999), Physics Reports, {\bf 314}, 575

  \bibitem{RH99} Rhoads, J.E. (1999), \apj{}, submitted; and
                 astro-ph/9903399

  \bibitem{SP99} Sari, R., Piran, T. and Halpern, J.P. (1999),
                 \apj{519}, L17

  \bibitem{Sh79} Shapiro, P.R. (1979), \apj{233}, 831

  \bibitem{Wo93} Woosley, S.E. (1993), \apj{405}, 273

\end{thebibliography}
\end{document}